\documentclass[11pt]{article}
\RequirePackage[OT1]{fontenc}
\usepackage{amsthm,amsmath,amssymb,amsfonts,bm,enumerate,xcolor,graphicx,natbib,color,url}
\RequirePackage[colorlinks,citecolor=blue,urlcolor=blue]{hyperref}
\usepackage{hyperref}
\usepackage{subcaption} 
\usepackage{ulem}
\usepackage{pifont}
\usepackage{textcmds}
\usepackage{booktabs}
\usepackage{multirow}
\usepackage{booktabs}
\usepackage{adjustbox}
\usepackage[usestackEOL]{stackengine}
\usepackage[toc,page]{appendix} 
\usepackage{algorithm,algorithmic}
\usepackage{lipsum}
\usepackage{setspace}

\usepackage{setspace}
\allowdisplaybreaks
\usepackage{bbm}

\DeclareMathOperator*{\argmin}{arg\,min}

\newcommand{\blind}{1}

\begin{document}
\thispagestyle{empty}
\baselineskip=28pt
\vskip 5mm

\renewcommand{\thefootnote}{\fnsymbol{footnote}}

\begin{center} 
{\Large{\bf Extreme Quantile Regression with Deep Learning}}\footnotemark[2]
\end{center}

\baselineskip=12pt

\vskip 5mm

\renewcommand{\thefootnote}{\arabic{footnote}}

\if1\blind
{
\begin{center}
\large
Jordan Richards$^1$ and Rapha\"el Huser$^2$
\end{center}
\renewcommand{\thefootnote}{\fnsymbol{footnote}} \footnotetext[2]{Extract of Chapter 21 (first 5 pages), to be published in the \emph{Handbook of Statistics of Extremes}, eds.\ Miguel de Carvalho, Rapha\"el Huser, Philippe Naveau and Brian Reich}
\renewcommand{\thefootnote}{\arabic{footnote}} \footnotetext[1]{
\baselineskip=10pt  School of Mathematics, University of Edinburgh, Edinburgh, UK. E-mail: jordan.richards@ed.ac.uk}
\footnotetext[2]{
\baselineskip=10pt Statistics Program, Computer, Electrical and Mathematical Sciences and Engineering (CEMSE) Division, King Abdullah University of Science and Technology (KAUST), Thuwal 23955-6900, Saudi Arabia. E-mail: raphael.huser@kaust.edu.sa}
\fi

\baselineskip=26pt
\vskip 2mm
\centerline{\today}
\vskip 4mm


\baselineskip=14pt

\section{Motivation}\label{sec:intro}

Estimation of extreme conditional quantiles is often required for risk assessment of natural hazards in climate and geo-environmental sciences and for quantitative risk management in statistical finance, econometrics, and actuarial sciences. For example, \cite{cooley2007bayesian} estimate high precipitation return levels (i.e., high marginal quantiles) over a region of Colorado using a Bayesian hierarchical model that incorporates geographical and climatic covariates, and \cite{cannon2010flexible,gardes2010conditional,velthoen2019improving} build flexible non-stationary models for hydroclimatological extremes; in the same vein, \cite{johannesson2022approximate} estimate spatio-temporal return levels of annual peak river flows to assess flood risk over the UK, while \cite{pasche2022neural} estimate high quantiles of river discharges to forecast flood risk in the Swiss Aare catchment; \cite{chavez2005generalized} build smooth covariate-dependent extreme-value models for low temperatures in the Swiss Alps, and \cite{zhong2022modeling} fit spatial extremes models to assess the spatio-temporal variability of extreme European heatwaves, in terms of their intensity, frequency, and spatial extent; \cite{richards2022regression,richards2023insights,koh2023spatiotemporal} build models to predict wildfire occurrences and extreme sizes over the US and the Mediterranean basin, which they then exploit to compute compound hazard estimates and assess climate change impacts; similarly, \cite{yadav2023joint} assess landslide hazard by fitting marked point process models designed to accurately capture both moderate and high quantiles of the landslide size distribution as a function of fixed covariates effects and unobserved random effects. As with environmental applications, the estimation of high quantiles (and extensions thereof) also plays a key role in the assessment of tail risk in finance \cite{chavez2018extreme,daouia2019extreme,daouia2023inference}, where the quantiles are often referred to as ``Value-at-Risk'' (VaR), and in insurance, for assessing risk associated with extreme claims and setting insurance premiums \cite{daouia2022extremile,daouia2023optimal}. Further applications and case studies are discussed in Part VI of this handbook (see, e.g., Chapters~25--27 for applications to natural hazards and Chapters~28--29 for applications to finance and the insurance industry).

In most of the above real data examples, interest lies in estimation of high (conditional) quantiles that exceed any past observations. Therefore, to extrapolate further into the tail, it is crucial to use a statistical framework that is well-adapted and especially designed for this purpose, and here extreme-value theory plays a key role. This chapter, which can be read as a follow-up of Chapters~6 and 20, precisely details how extreme quantile regression may be performed using theoretically-justified models, and how modern deep learning approaches can be harnessed in this context to enhance the model's performance in complex high-dimensional settings. While a variety of extreme-value techniques have been advocated for this purpose, we here focus on \emph{parametric} quantile regression, exploiting well-established extreme-value distributions. Similarly, while several machine learning approaches have been combined with regression in the literature, we argue that deep learning methods based on neural networks with suitable architectures are ideally placed to tackle truly high-dimensional regression problems (where the number of covariates is large). The power of deep learning combined with the rigor of theoretically-justified extreme-value methods opens the door to efficient extreme quantile regression, in cases where both the number of covariates and the quantile level of interest can be simultaneously ``extreme''.

The rest of the chapter is organised as follows: Section~\ref{sec:eqr} gives a gentle introduction to parametric extreme quantile regression, while Section~3 introduces deep regression and basic neural networks, and details how deep extreme quantile regression models may be built and trained, with some final comments on software availability. Section~4 discusses a small simulation study to compare various statistical and machine learning competitors, while Section~5 illustrates the methods by application to the estimation of European precipitation return levels. Section~6 finally discusses further important topics, and concludes with some perspectives for future research.

\section{An Introduction to Extreme Quantile Regression}\label{sec:eqr}
After introducing the general quantile regression setting in Section~\ref{sec:setting}, we briefly explain in Section~\ref{sec:eqr-classicmethods} why classical non-parametric quantile regression methods fail when interest lies in the extreme tails; we then review parametric extreme quantile regression in Section~2.3 and finally make the case for using deep learning in that context in Section~2.4. For a more detailed review of classical parametric regression models for extreme values, see Chapter~6.

\subsection{General Setting}\label{sec:setting}
We consider the setting where $Y\in\mathbb{R}$ is a continuous random variable of interest, referred throughout as the \textbf{response} variable, and which has distribution function $F_Y(y)=\text{P}(Y \leq y)$. For $\tau\in(0,1)$, the $\tau$-quantile of $Y$ is $Q(\tau):=F^{-1}_Y(\tau)=\inf\{y : F_Y(y) \geq \tau\}$, where $F^{-1}_Y(\cdot)$ denotes the inverse of $F_Y(\cdot)$. We now introduce a $q$-dimensional vector of continuous-valued covariates or ``predictors'', denoted by $\mathbf{X}=(X_1,\dots,X_q)^T\in\mathbb{R}^q$. In a \textit{regression} setting, we seek to relate covariates $\mathbf{X}$ to $Y$ through the conditional distribution function 
$$F_{Y|\mathbf{X}}(y\;|\;\mathbf{x})=\text{P}(Y \leq y\;|\;\mathbf{X}=\mathbf{x}).$$ 

\textit{Quantile regression} describes the act of relating $\mathbf{X}$ to a specific quantile of $Y$, i.e., estimating the conditional $\tau$-quantile of $Y$, given observed covariates $\mathbf{X}=\mathbf{x}$, defined by
\begin{equation}\label{eq:cond_quant}
Q_\mathbf{x}(\tau):=F^{-1}_{Y|\mathbf{X}}(\tau\;|\;\mathbf{x})=\inf\{y : F_{Y|\mathbf{X}}(y\:|\;\mathbf{x}) \geq \tau\}.
\end{equation}
Modelling of \eqref{eq:cond_quant} requires specification of the functional form for $Q_\mathbf{x}(\tau)$ (viewed as a function of the input $\mathbf{x}$ for fixed $\tau$). 
Popular approaches specify \eqref{eq:cond_quant} to be a linear or additive function (of the components of $\mathbf{x}$), up to some non-linear transformation via a link function. For a general overview of quantile regression, see \cite{koenker2005quantile} or the more recent review by \cite{koenker2017handbook}.

\textit{Extreme quantile regression} refers to scenarios where $\tau$ is either very close to one, such that $Q_\mathbf{x}(\tau)$ is typically larger than all observations of $Y\;|\;(\mathbf{X}=\mathbf{x})$ available for inference, or $\tau$ is close to zero, where the converse holds; in short, $Q_\mathbf{x}(\tau)$ is often beyond the range of the observed data. Without loss of generality, we focus on the case where $\tau$ is close to one, and thus consider estimation of the upper-tails of $Y\;|\;(\mathbf{X}=\mathbf{x})$ only. Formally, \cite{chernozhukov2017extremal} define extreme quantile regression by considering estimation of $Q_\mathbf{x}(\tau)$ for a sequence of quantile levels. Let $n$ denote the sample size of observation pairs $\{(y_1,\mathbf{x}_1),\dots,(y_n,\mathbf{x}_n)\},$ and let $\{\tau_n\}_{n=1}^\infty$ be a sequence of quantile levels which are dependent on $n$. Extreme conditional quantiles are those for which ${\tau_n\rightarrow 1}$ and $n(1-\tau_n) \rightarrow k \in[0,\infty)$ as $n\rightarrow \infty$, where $n(1-\tau_n)$ denotes the expected number of exceedances of $y_i$ above $Q_{\mathbf{x}_i}(\tau_n)$. Hence, there are a finite (or possibly zero) number $k$ of exceedances of $Q_{\mathbf{x}_i}(\tau_n)$ in the sample as $n\rightarrow \infty$. Such data scarcity causes standard non-parametric methods, i.e., those not adapted for the tails, to perform poorly when $\tau_n$ is close to one. This issue is further discussed in the next section.

\subsection{Why Not Use Classical Non-Parametric Methods?}\label{sec:eqr-classicmethods}

In classical least squares regression, the conditional expectation  $\text{E}[Y \: | \; \mathbf{X}=\mathbf{x}]$ is found by minimising a squared error prediction loss. The analogue in quantile regression is the minimisation problem 
\begin{equation}\label{eq:quant_loss}
Q_\mathbf{x}(\tau):=\argmin_\beta \text{E}[\rho_\tau(Y-\beta)\mid\mathbf{X}=\mathbf{x}],
\end{equation}
where $\rho_{\tau}(u):=u(\tau-\mathbbm{1}\{u < 0\})$ is the asymmetric absolute deviation function \cite{fox1964admissibility}, more commonly termed the ``pinball'' \cite{koenker1978regression} or ``quantile check'' loss (see \cite{gneiting2011quantiles} for alternative nomenclature). In practice, one models $Q_\mathbf{x}(\cdot)$ by constructing a set of quantile functions that map $\mathbf{x}$ to some real value; we denote such a set of functions by $\mathcal{Q}$, with feasible non-decreasing quantile functions as $q_\tau(\cdot) \in \mathcal{Q}$. In classical non-parametric quantile regression, estimates $\widehat{Q}_\mathbf{x}(\tau)$ of the conditional $\tau$-quantile of $Y\:|\;(\mathbf{X}=\mathbf{x})$ are then yielded by minimising empirical estimates of \eqref{eq:quant_loss} over observation pairs, that is, 
\begin{equation}\label{eq:emp_quant_loss}
\widehat{Q}_\mathbf{x}(\tau):=\argmin\limits_{q_\tau \in \mathcal{Q}} \frac{1}{n}\sum^n_{i=1}\rho_\tau\{y_i-q_{\tau}(\mathbf{x}_i)\}.
\end{equation}
The method is referred to as ``non-parametric'' in the sense that it does not assume any particular parametric probability model for $Y\:|\;(\mathbf{X}=\mathbf{x})$, since the loss function $\rho_\tau$ is quantile-specific and does not impose any structure across the different values of $\tau$. 
Choices of quantile functions $q_\tau(\cdot) \in\mathcal{Q}$ can, however, be parametric, semi-parametric, or non-parametric themselves (as a function of $\mathbf{x}$), and many options have been proposed in the literature. The de facto standard is the 
linear model, popularised by \cite{koenker1978regression}. 
More flexible alternatives which utilise additive functions, such as splines, have also been proposed \cite{koenker1994quantile,fasiolo2021fast}.  Due to an ever-increasing supply of high-dimensional and massive datasets, more recent proposals have adopted machine learning algorithms \cite{zhong2023neural}, which benefit from the flexibility and computational scalability not offered by linear or additive models, and represent $q_\tau(\cdot)$ as a highly non-linear, non-additive 
function. 

However, when interest lies in estimating extreme quantiles and extrapolating to values of $\tau$ that go much beyond $1-1/n$, classical non-parametric methods that rely on solving the minimisation problem~\eqref{eq:emp_quant_loss} fail miserably, whatever the generality of $\mathcal{Q}$ and the flexibility or functional form of $q_\tau(\cdot)$. To illustrate this well-known problem, we consider a simple unconditional setting without covariates, and simulate random samples $Y_1,\ldots,Y_n\sim F_Y$ of size ${n=1000}$, independently from four distributions $F_Y$ with increasing tail heaviness. These are 
the Normal distribution with zero mean and unit variance (${\rm N}(0,1)$), the Gamma distribution with scale~$1/4$ and shape~$4$ (${\rm Gamma}(0.25,4)$), the log-normal distribution with zero log-mean and unit log-variance (${\rm logN}(0,1)$), and the Fr\'echet distribution with shape~$3$ ({\rm Fr\'echet}$(3)$). For each random sample, we compute empirical $\tau$-quantiles with $\tau\in[0.99,0.99999]$; that is, the exceedance probability ranges from $10^{-2}$ (with about $10$ exceedances among the $n$ observations) to $10^{-5}$ (with no exceedances most of the time). When $\tau$ is larger than $1-1/n$, the empirical $\tau$-quantile will always be estimated as the sample maximum, $\max(Y_1,\ldots,Y_n)$, which is independent of $\tau$. Therefore, the bias of the empirical quantile estimator is expected to grow with $\tau$. 

Figure~\ref{GPDvsEmpirQuantiles} shows the results of this simulation study. 
\begin{figure}[t!]
    \centering
    \includegraphics[width=0.8\linewidth]{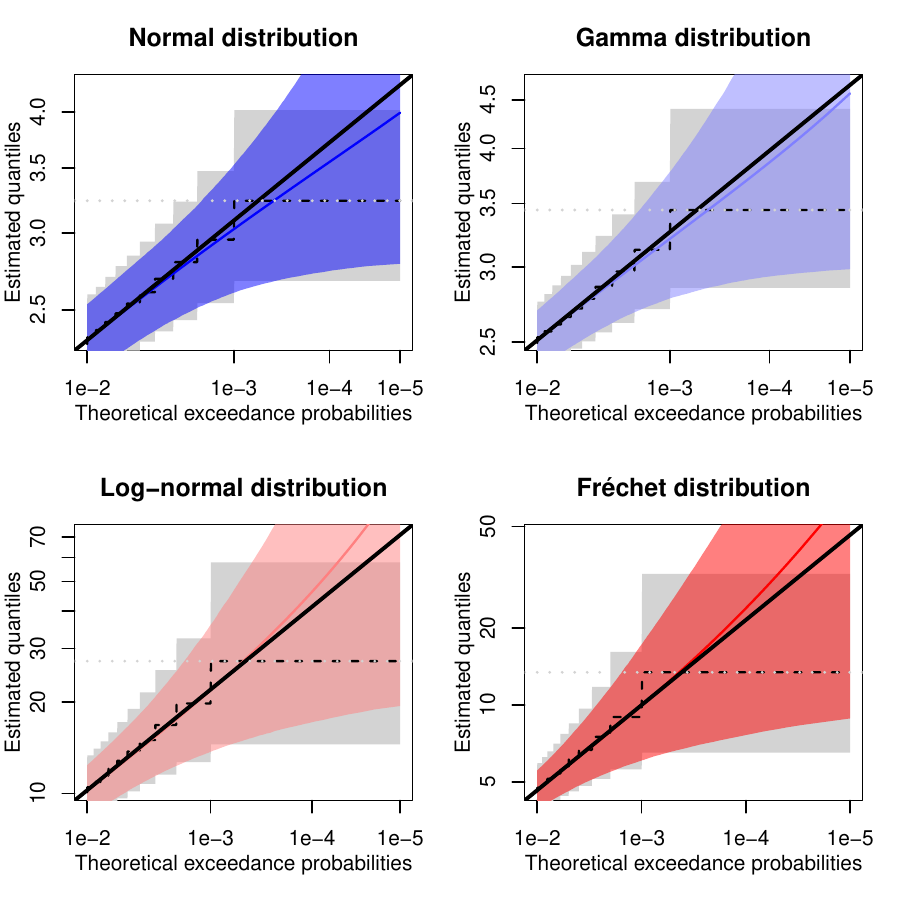}
    \caption{Empirical and GP-based $\tau$-quantile estimates plotted against $1-\tau$  with $\tau\in[0.99,0.99999]$. Estimates are from random samples of size $n=1000$, in four cases: ${\rm N}(0,1)$ (top left), ${\rm Gamma}(0.25,4)$ (top right), ${\rm logN}(0,1)$ (bottom left), and {\rm Fr\'echet}$(3)$ (bottom right). From $10^4$ simulation replicates, empirical quantiles are displayed through their average (dashed black) and their $95\%$ variability (grey shaded envelope), while GP-based estimates are displayed through their average (solid colored) and their $95\%$ variability (colored shaded envelope). The thick diagonal line corresponds to a perfect fit. The grey dotted horizontal line represents the value of $\text{E}[\max(Y_1,\ldots,Y_n)]$ in each case.}
    \label{GPDvsEmpirQuantiles}
\end{figure}
The dashed black line shows the empirical $\tau$-quantile, averaged across $10^4$ simulations, while the grey shaded area shows its sampling variability obtained through the $2.5\%$ and $97.5\%$ quantiles across the simulated replicates. In all cases, the empirical quantile estimates are reasonable up to $\tau=1-1/n=1-10^{-3}$ but they quickly deteriorate as $\tau$ increases further. In fact, the estimator becomes increasingly biased since it is constant beyond that probability level. While alternative smoother empirical estimators are available, they are always expressed as convex combinations of order statistics, and thus will display similar drawbacks. These results illustrate clearly the crucial need for extreme quantile regression methods that rely on extreme-value theory for tail extrapolation. To illustrate the benefit of ``parametric'' extreme-value regression techniques in this context, we also fit, for each simulated dataset, the generalised Pareto (GP) distribution to threshold exceedances above the empirical $95\%$ quantile and then used the fitted model to perform tail extrapolation; see Chapters~2--3 (as well as Section~2.3 or \cite{davison2015statistics}) for further details on the definition and justification of the GP distribution, as well as other extreme-value models. The results from the parametric GP fits are displayed in Figure~\ref{GPDvsEmpirQuantiles} using solid colored curves (averaged across $10^4$ simulations) and colored shaded areas (delimiting the $2.5\%$ and $97.5\%$ quantiles of the estimates across the $10^4$ simulated replicates). While the uncertainty of GP-based quantile estimates grows quite fast as $\tau$ increases, estimates are typically more accurate than non-parametric estimates and display a much smaller bias for large $\tau$.

\baselineskip 12pt
\bibliographystyle{CUP}
\bibliography{bibtex_example}

\end{document}